\documentclass[sigconf]{nimeart}

\usepackage[dvipsnames]{xcolor}
\usepackage{array}
\usepackage{multirow}
\usepackage{enumitem}

\usepackage{xurl}

\usepackage{CJKutf8}

\AtBeginDocument{%
  }

\setcopyright{cc}
\copyrightyear{2026}
\acmYear{2026}
\acmDOI{}

\acmConference[NIME '26]{International Conference on New Interfaces for Musical Expression}{June 23--26,
  2026}{London, UK}
\acmISBN{}

\begin{document}

\title{Understanding Listener Perceptions of AI and Human-Composed Music in Emotional Applications}

\author{Kimaya Lecamwasam}
\affiliation{%
  \institution{MIT Media Lab}
  \city{Cambridge}
  \country{USA}}
\email{klecamwa@media.mit.edu}

\author{Tishya Ray Chaudhuri}
\affiliation{%
  \institution{Myndstream}
  \city{London}
  \country{UK}
}

\renewcommand{\shortauthors}{Lecamwasam and Ray Chaudhuri}

\begin{abstract}
  Designing music-based affective technologies requires understanding of how perceptions of AI versus human authorship shape trust and authenticity. We investigate how listener perception of AI-generated versus human-composed music affects emotional resonance and regulation. Drawing on affective computing and human-computer interaction frameworks, participants listened to AI- and human-composed music across labeling conditions (\textcolor{ForestGreen}{Correct}, \textcolor{red}{Incorrect}, or \textcolor{orange}{Unlabeled}) and emotion cases (\textcolor{blue}{Calm} and \textcolor{violet}{Upbeat}). Participants rated preference, efficacy of target emotion elicitation, and emotional impact. Results showed participants found human-composed music more effective in eliciting their target affective states and linked humanness to imperfection, flow, and “soul,” underscoring authenticity as central to appraisal and ultimately leading to design implications relevant to music-based HCI. These findings challenge the assumption that preference alone defines system success, highlighting design implications for affective and wellness technologies that foreground authenticity, transparency, and human creativity.
\end{abstract}


\begin{teaserfigure}
  \begin{center}
  \includegraphics[width=\textwidth]{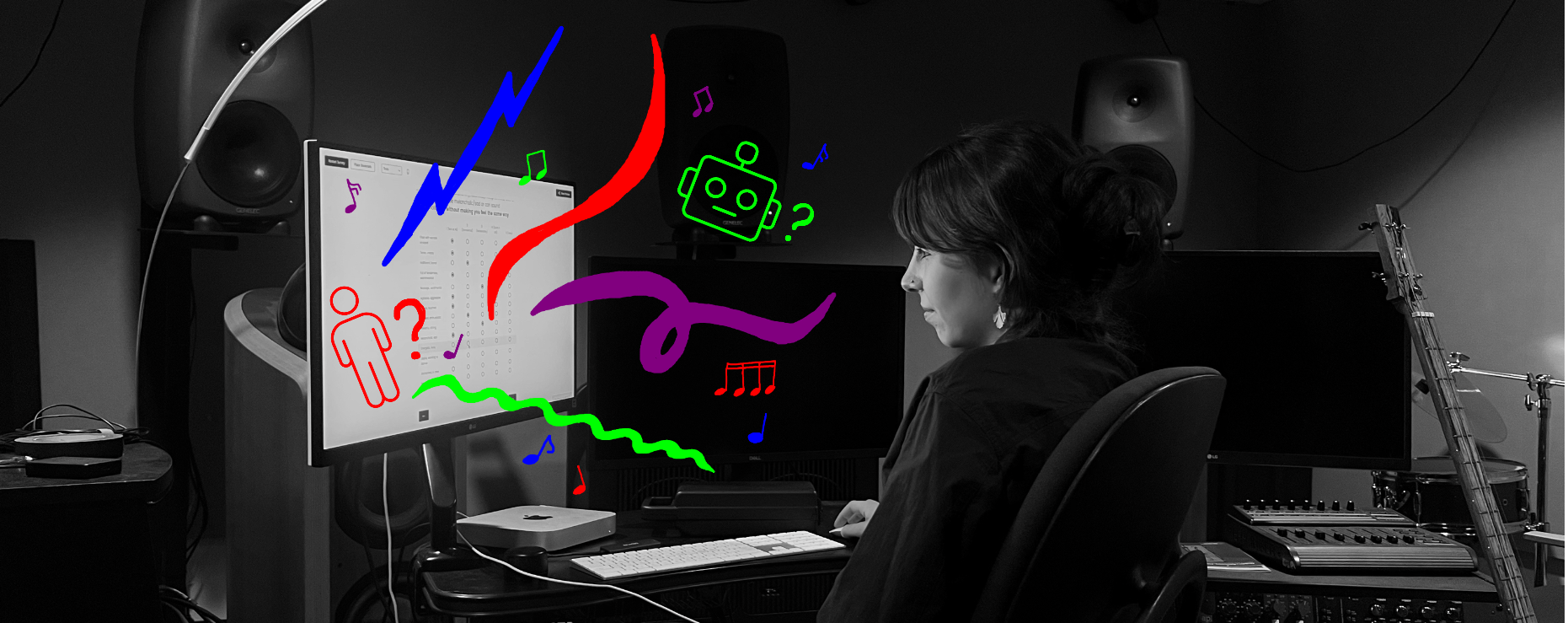}
  \Description{Black and white image of a woman looking at a computer screen displaying the survey from the study protocol, with hand-drawn colorful squiggles and music notes, as well as the outlines of a human figure and a robot head, emerging from the screen.}
  \end{center}
  \label{fig:teaser}
\end{teaserfigure}

\maketitle

\section{Introduction} 
Generative AI is changing the shape of music creation. Despite its increasing integration, it is unclear whether the emotional impact of AI-generated music matches that of human-composed. To assess this, we conducted a study ($N=152$) exploring how listeners classify and perceive AI and human music (Fig.~\ref{fig:summary_fig}). Participants listened to four one-minute-long, instrumental-only audio clips (two human and two AI) from two emotion cases: ``Calming/Soothing'' (\textcolor{blue}{Calm}) and ``Upbeat/Invigorating'' (\textcolor{violet}{Upbeat}). Participants were split into three groups: \textcolor{orange}{Unlabeled}, \textcolor{ForestGreen}{Correct}ly labeled, and \textcolor{red}{Incorrect}ly labeled (Figure~\ref{fig:summary_fig}-I). All participants completed demographics and validated pre/post-task questionnaires assessing state anxiety~\cite{spielberger1971state} (exploratory measure) and musical emotional resonance~\cite{coutinho2017introducing}, indicated which songs they preferred and thought most effectively conveyed the target emotion, and provided reasoning. This study was preregistered prior to data collection on the Open Science Framework~\cite{lecamwasam2024prereg}. We hypothesized\footnote{Hypotheses condensed for clarity.}: 

\begin{enumerate}[label=\textbf{H\arabic*}, leftmargin=*, align=left]
    \item Without origin labels, participants will prefer human-composed music and find it more effective.
        \label{h:unlabeled_general}
    \begin{enumerate}[label=\textbf{H\arabic{enumi}.\alph*}, leftmargin=*, align=left]
    \item For \textcolor{violet}{Upbeat} music, unlabeled human-composed music will yield higher GEMIAC scores in high arousal/high valence emotion classes~\cite{mesz2023marble}. 
        \label{h:unlabeled_cs}
    \item For \textcolor{blue}{Calm} music, unlabeled human-composed music will yield higher GEMIAC scores in low arousal/high valence emotion classes~\cite{mesz2023marble}.
        \label{h:unlabeled_ui}
    \end{enumerate}
    \item Participants will prefer, and find more effective, songs labeled human, regardless of actual origin. 
        \label{h:labeled_general}
    \begin{enumerate}[label=\textbf{H\arabic{enumi}.\alph*}, leftmargin=*, align=left]
    \item For \textcolor{violet}{Upbeat} music, songs labeled human, regardless of actual origin, will yield higher GEMIAC scores in high arousal/high valence emotion classes~\cite{mesz2023marble}.
        \label{h:labeled_cs}
    \item For \textcolor{blue}{Calm} music, songs labeled human, regardless of actual origin, will yield higher GEMIAC scores in low arousal/high valence emotion classes~\cite{mesz2023marble}. 
        \label{h:labeled_ui}
    \end{enumerate}
    \item Interaction with musical stimuli will impact momentary anxiety measures, though this requires further study.
        \label{h:stai}
\end{enumerate}

These hypotheses are consistent with previous work showing that musicians communicate discrete emotions through acoustic cue patterns (e.g. tempo, dynamics, and articulation) and that expectation management and expressive microtiming shape listeners’ affective responses~\cite{juslin2003communication,huron2008sweet,senn2017rhythmic,gabrielsson1996emotional}. By contrast, text-to-music systems have been limited by a lack of ability to ``precisely control musical features so that the resulting music exhibits the desired affect''~\cite{dash2024ai} due to interdependencies, dataset limitations, and lack of transparency~\cite{dash2024ai,briot2020deep,carnovalini2020computational}. Our findings suggest participants tend to find human music more effective in eliciting target emotions but skew towards preference for AI. Additionally, though GEMIAC \textit{category} representation was not impacted by origin source, GEMIAC \textit{scores} were significantly higher for human music in specific cases. Qualitative data also revealed persistent associations between human composition and emotional authenticity.

\section{Related Work}
\subsection{Emotion-Sensitive AI and Ethics}
Given music's well-established impact on human perception, cognition, and emotion~\cite{north2004uses,sarkamo2013music}, there is growing interest in the use of \textit{generative} music in similar contexts, including health interventions~\cite{shen2024first,rahman2021towards}, adaptive soundtracks~\cite{hutchings2019adaptive,ibanez2021using}, and personalized recommendations~\cite{sharma2024feel,sharma2021emotion}. This interest reflects broader trends in AI research towards developing systems that better understand and respond to human emotions. However, ethical considerations remain, including debates over the universality of emotional expression and the implications culture-dependent and nature-nurture nuance hold for emotion-centered algorithms~\cite{ekman1970universal,jack2012facial,scherer2011eye}. In emotion regulation contexts, clinical algorithms must (1) minimize bias, (2) promote user autonomy and confidentiality, and (3) be clinically effective~\cite{straw2021ethical,fiske2019your}. We seek to contribute to this larger conversation by investigating the musical features that are highlighted as ``human'' by human listeners, to assess how authenticity, emotional resonance, and cultural context shape human-computer interactions.

\subsection{Ethics and Generative Music}
Current perspectives on AI in music creation/production raise questions regarding copyright law~\cite{sturm2019artificial}, training bias~\cite{mehta2024missing}, and practical application~\cite{barnett2023ethical}. In a multi-subject case study of working musicians from a variety of genres and disciplines, researchers highlighted musicians' focus on the threat AI poses to employment and artistic integrity, in part due to concerns about a lack of both transparency and creative control within AI tools~\cite{newman2023human}. Recent advances in music generation have intensified concerns. SunoAI~\cite{suhailudheen2025suno}, Udio~\cite{udio2024}, and Lyria~\cite{team2025live} have demonstrated unprecedented ability to generate music across genres, as evidenced by the 2025 Deezer and Ipsos survey ($N=9000$) which revealed 97\% of participants found AI-generated music perceptually indistinguishable from human-composed~\cite{wendel2025deezer}. Results such as these raise urgent questions about the future of artistic integrity, creator attribution, and economic sustainability in music creation~\cite{rahman2024sonics}. Furthermore, many existing systems rely heavily on Western tonal music, reducing cross-cultural applicability~\cite{athanasopoulos2021harmonic}. This parallels current emotion-sensitive systems' tendencies to oversimplify emotion states, mapping them to culturally limited dimensions, thereby failing to capture the complexity and nuance of human emotion~\cite{huang2023beyond,mehta2024missing}.

\subsection{Existing Comparisons of AI-Generated and Human-Composed Music}
Tigre Moura's and Maw's (2021) participants reported negative attitudes and low purchase intentions for AI music, though awareness of AI did not impact actual music evaluations, suggesting a disconnect between opinions and perceptual experiences~\cite{tigre2021artificial}. Hong et al. (2022) found that, while anthropomorphizing generative AI systems did not affect music quality scores, accepting AI as a ``musician'' led to higher ratings, suggesting that role perception drives evaluations more than sonic attributes~\cite{hong2022human}. Zenieris (2023) demonstrated this labeling effect directly. Participants who knew whether songs were AI-generated or human-composed rated human music 80\% higher, while uninformed participants preferred AI 66.7\% of the time~\cite{zenieris2023perception}. Sun et al. (2023) found that anthropomorphic features enhanced AI music's perceived competence and warmth~\cite{sun2024would}, while Liu (2025) found that emotional attachment drove continued use of AI music platforms~\cite{liu2025impact}. Chu et al. (2022) compared music generation models, linking melodiousness to highest satisfaction, though token representation methods and model characteristics also had significant impact~\cite{chu2022empirical}. Fernando et al. (2024)'s review of studies of emotional responses to AI music found that, while some systems show promise, ongoing listener skepticism calls for more research on emotional authenticity~\cite{fernando2024assessment}. 

However, critical gaps remain, particularly related to the impact of AI on emotion regulation. While Hong et al. and Zenieris showed labeling that affects preference, neither examined whether this also impacts emotional efficacy~\cite{hong2022human,zenieris2023perception}. Additionally, while Tigre Moura and Maw and Zenieris examined responses when labels were experimenter-provided, few studies have systematically assessed how listeners' identification accuracy of unlabeled AI-versus-human music affects subsequent judgments~\cite{tigre2021artificial,zenieris2023perception}. We aim to address these gaps by investigating: (1) systematically distinguishing preference (aesthetic appeal) from efficacy (perceived emotional/therapeutic effectiveness) using within-participant comparisons and validated emotional measures; (2) labeling accuracy in an \textcolor{orange}{Unlabeled} condition where participants identify music origin; (3) emotion-specific effects for \textcolor{blue}{Calm} versus \textcolor{violet}{Upbeat} music; (4) directional preference-efficacy consistency to reveal whether these constructs align or diverge; and (5) how listener mislabeling rates vary and whether this predicts preference. 

\begin{figure*}[]
 \centering
\includegraphics[width=0.8\textwidth]{M_StudyOverview_Revised_II.png}
 \caption{(\textit{I}) -- Participants completed demographics/STAI-S questionnaires, and were randomly assigned to one experimental group: (A) \textcolor{ForestGreen}{Correct}ly labeled AI-generated and human-composed music; (B) \textcolor{red}{Incorrect}ly labeled music (AI labeled as human and vice versa); and (C) \textcolor{orange}{Unlabeled} music, where participants labeled which songs they thought were human and which they thought were AI. They completed the STAI-S again at the end of the study. (\textit{II}) -- All listened to four songs: two \textcolor{blue}{Calm} (one AI, one human) and two \textcolor{violet}{Upbeat} (one AI, one human) in randomized order. Following each song, participants completed a labeling/preference task. Figure created in https://BioRender.com}
 \label{fig:summary_fig}
 \Description{Part I of the figure, on the far left, depicts the three experimental conditions used in the study. participants first completed demographic questions and a baseline state anxiety measure (STAI-S), shown in a grey box. They were then assigned to one of three experimental conditions. In the Correct Experimental Group (C), shown in green, participants listened to both human-composed and AI-generated music, each with two emotional cases (Calming/Soothing and Upbeat/Invigorating), and the pieces were labeled accurately according to their origin. In the Incorrect Experimental Group (I), shown in red, the same two types of music and emotional cases were presented, but the authorship labels were intentionally swapped so that human-composed pieces were labeled as AI and AI-generated pieces were labeled as human. Finally, in the Unlabeled Experimental Group (U), shown in orange, participants also heard both human- and AI-composed music across the two emotion conditions, but no authorship labels were provided. The layout emphasizes that all groups encountered the same Calming/Soothing and Upbeat/Invigorating music, with only the labeling strategy differing across conditions. After completing this phase, participants again reported their state anxiety (STAI-S), shown in a grey box. Part II of the figure, on the far right, shows that within each condition, participants listened to both Calming/Soothing and Upbeat/Invigorating music excerpts.  On the right, the figure shows that in the second phase, participants heard both human- and AI-composed music paired with the GEMIAC emotional response survey. The order of presentation was randomized, and after each excerpt participants provided labeling judgments and preference ratings. This layout illustrates the sequencing of measures and experimental manipulations across the study.}
 \end{figure*}

\section{Methods}
This study was conducted on Prolific via Qualtrics survey ($N=152$) in September 2024. Participants listened to \textcolor{blue}{Calm} and \textcolor{violet}{Upbeat} music, chosen from the most requested use-cases identified by Myndstream, a wellness music company\footnote{More information about Myndstream can be found at: \url{myndstream.com}}. There were two one-minute-long songs within both cases, one generated by SunoAI and the other composed in-house by Myndstream using digital instruments (four total). SunoAI and the producer received the same two, 200-word prompts written by Myndstream’s Head of Music based on briefs given to human composers to compose for the two emotion cases\footnote{Please see Supplementary Material~\ref{appendix:music} for descriptions of and links to all songs used in the protocol.}:
\begin{itemize}
    \item \textcolor{blue}{Calm}: ``Compose a relaxing instrumental piece to eliminate anxiety. Use soothing melodies, gentle harmonies, and calm tempo. Include soft piano, ambient synths and ethereal effects to evoke peace and relaxation.''
    \item \textcolor{violet}{Upbeat}: ``Compose an invigorating instrumental piece to boost mood. Use uplifting melodies, dynamic rhythms, and a lively tempo. Include electronic drums, vibrant synths, and ethereal effects to evoke joy.''
\end{itemize}
Both the producer and the researcher who generated the music did not have access to the other case to prevent bias. The AI-generated song was chosen after one round of generation without post-production. 

Participants completed a demographic questionnaire (music training/preference, opinions on AI in music, experience with generative AI, etc.) and the State Anxiety Scale from the State-Trait Anxiety Inventory (STAI-S)~\cite{spielberger1971state}, due to its sensitivity to short-term interventions, though STAI-S data were purely exploratory. Participants were randomly assigned to one experimental group: (1) \textcolor{ForestGreen}{Correct}ly labeled AI-generated and human-composed music; (2) \textcolor{red}{Incorrect}ly labeled music, where AI-generated music was labeled as human-composed and vice versa; and (3) \textcolor{orange}{Unlabeled} music, where participants labeled which music they thought was AI-generated and which they thought was human-composed (Fig.~\ref{fig:summary_fig}-I). At the end of the protocol, \textcolor{red}{Incorrect} participants were informed of the mislabeling and were asked to reflect. All participants received a disclaimer about potential mislabeling to meet ethical guidelines (MIT COUHES protocol E-6149).

Participants were randomly assigned to begin with either \textcolor{blue}{Calm} or \textcolor{violet}{Upbeat} music as part of a fully factorial design (Fig.~\ref{fig:summary_fig}-II). Within each emotion block, all participants listened to one song at a time (randomly assigned to either AI or human first) and, immediately after listening to each song, completed the Geneva Music-Induced Affect Checklist (GEMIAC) to assess the emotional resonance of each piece~\cite{coutinho2017introducing}. The GEMIAC survey builds upon the Geneva Music Emotion Scale (GEMS) by incorporating a broader range of genres and positive/negative emotions~\cite{coutinho2017introducing}. We used the intensity rating\footnote{Please see Supplementary Material~\ref{appendix:gemiac} for a mockup of the GEMIAC intensity scale.} due to the short duration and ``relatively homogeneous emotional tonality'' of the stimuli, following Coutinho's and Scherer's recommendation~\cite{coutinho2017introducing}.

Participants also compared each song pair, reporting song which they preferred and which more effectively conveyed the target emotion (``efficacy'') (Fig.~\ref{fig:summary_fig}-II). \textcolor{orange}{Unlabeled} participants labeled which song they believed was human and which was AI. All participants reflected on the musical features that influenced their choices. Finally, participants completed the STAI-S again. This study was randomized and included repeated measures, using a mixed design to compare both between- and within-subject data. We embedded two attention checks into the survey.

\subsection{Data Analysis}
\subsubsection{Quantitative Analysis}
We analyzed preference and efficacy ratings using multinomial logistic regression (\texttt{nnet} R package~\cite{ripley2016package}), since both outcomes were categorical with three unordered levels (AI-Generated, Human-Composed, Neither/Unclear). Models included experimental group, emotion case, and their interaction, with Type III (Wald) chi-square tests for overall effects and Bonferroni-corrected pairwise comparisons via the \texttt{emmeans} package~\cite{emmeansR2025}. We assessed within-participant preference-efficacy agreement using generalized linear mixed-effects models (GLMMs) with binomial distributions (\texttt{lme4} package~\cite{bates2015package}), including random intercepts for participants. We conducted directional consistency analyses using Fisher's Exact Tests to determine whether AI-preferrers also rated AI as more effective, and compared consistency rates between AI- and human-preferrers. We assessed \textcolor{orange}{Unlabeled} group labeling accuracy by comparing participants' labels against actual music origin and used Fisher's Exact Tests to examine whether accuracy predicted preference/efficacy. Demographic effects were examined using GLMMs predicting (1) preference-efficacy agreement, (2) binomial AI preference, and (3) \textcolor{orange}{Unlabeled} participants labeling accuracy. Age ranges were converted to numeric midpoints. All models included experimental group and emotion as covariates. GEMIAC scores were analyzed using linear mixed-effects models to assess how age, prior experience with AI-generated music, general music experience, and opinions on AI in the music industry impacted scores, with fixed effects for experimental group, emotion condition, music origin, GEMIAC category, and their interactions and random intercepts for participants to account for repeated measures across observations. All analyses used $\alpha = 0.05$ with Bonferroni correction for multiple comparisons, conducted in R version 4.5.1~\cite{rcoreteam2021}. 

\subsubsection{Qualitative Analysis} 
Free responses were analyzed using Braun and Clarke's guidelines~\cite{clarke2017thematic}. We generated initial codes inductively, ranging from surface-level observations (e.g., ``mentions rhythm'', ``describes vocals'') to latent meanings (e.g., ``prefers human imperfection'', ``distrusts AI''). Themes were grouped by shared meaning: ``organic'' and ``alive” were clustered under \textit{Naturalness and Humanity}, while ``emotionally moving'' and ``soothing'' supported \textit{Emotional Resonance}. We examined internal coherence and refined themes to ensure distinction. We note that free response data were analyzed thematically by a single rater. Given the absence of inter-rater reliability assessment, these findings should be considered exploratory and interpreted with appropriate caution. Future work should incorporate multiple raters and formal reliability metrics to validate identified patterns.

\section{Results}
 \begin{figure*}[]
   \includegraphics[width=0.8\textwidth]{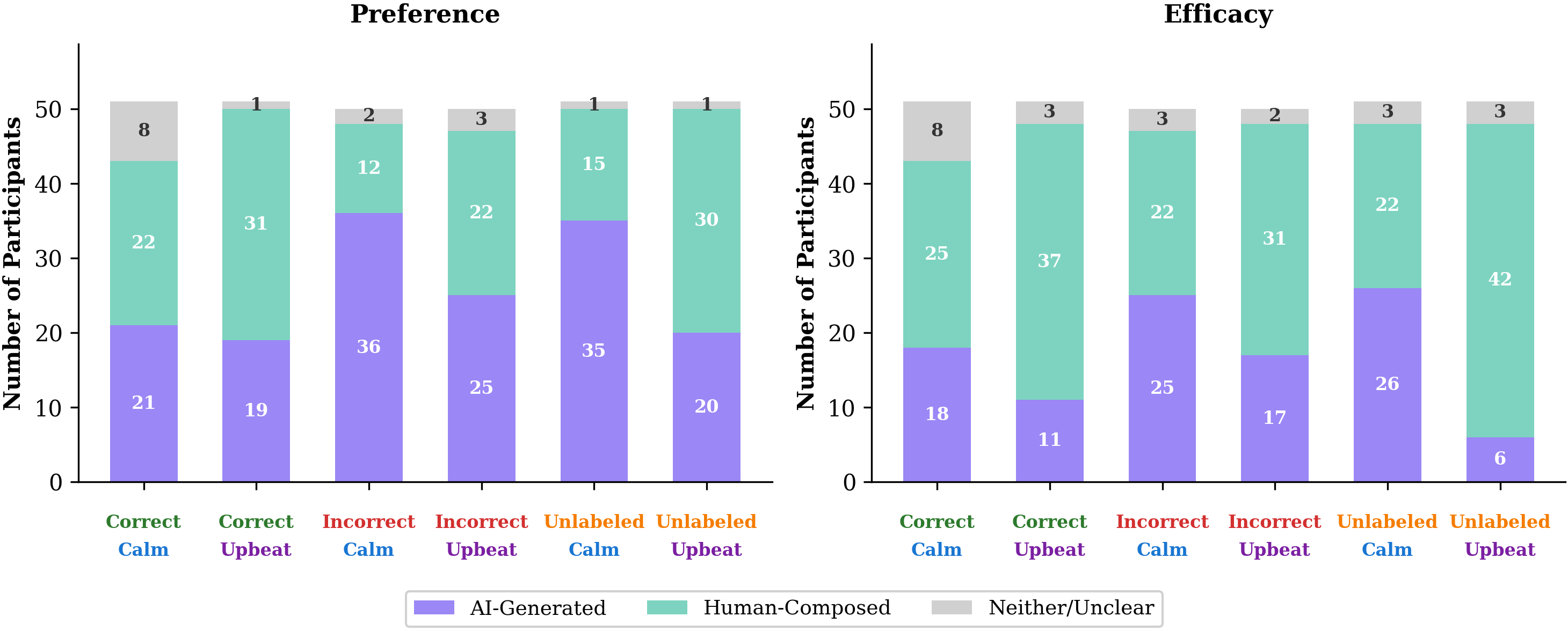}
   \caption{Stacked bar plots showing the number of participants in each experimental group (\textcolor{ForestGreen}{Correct}, \textcolor{red}{Incorrect}, \textcolor{orange}{Unlabeled}) who indicated (left) preference for and (right) perceived efficacy of AI-generated music, human-composed music, or neither/unclear, separated by emotion (\textcolor{blue}{Calm}, \textcolor{violet}{Upbeat}). Numbers indicate participant counts within each category.}
   \Description{This figure shows side by side stacked bar plots illustrating participant preference and efficacy counts across experimental groups and emotion conditions. AI-generated responses are shown as the purple fraction of the bar, human-composed responses are shown as blue-green, and neither/unclear responses are shown as grey. The x-axes shows paired emotion class and experimental group (e.g. ``correct calm'' or ``incorrect upbeat.'' Y-axes show the number of participants, ranging from 0 to 50.}
   \label{fig:pref_eff}
 \end{figure*}
\subsection{Preference, Efficacy, and Labeling}
Overall, participants rated human-composed music as more effective (58.9\%) than AI (33.9\%), while preferences were more balanced (51.2\% AI vs 43.1\% human across both emotions). Multinomial logistic regression revealed a significant difference between preference and efficacy ($\chi^2(2) = 19.03, p < 0.001$). Specifically, participants were 52\% less likely to rate AI music as effective compared to their preference ratings ($OR = 0.48, 95\% CI [0.36, 0.65], \\z = -4.25, p < 0.001$), indicating a substantial preference-efficacy disconnect (Hypotheses~\ref{h:unlabeled_general},~\ref{h:labeled_general}).

\subsubsection{Impact of Experimental Group} 
Experimental group significantly predicted music preference ($\chi^2(4) = 16.51, p = 0.002$). The \textcolor{red}{Incorrect} group (72\% AI preference for \textcolor{blue}{Calm}, 50\% for \textcolor{violet}{Upbeat}) showed significantly higher AI preference (though the music was labeled as human during the task) than the \textcolor{ForestGreen}{Correct} group (41\% AI for \textcolor{blue}{Calm}, 37\% for \textcolor{violet}{Upbeat}; $OR = 3.00, 95\% CI [1.21, 7.43], z = 2.42, p\_adj = 0.047$) (Hypotheses~\ref{h:labeled_cs},~\ref{h:labeled_ui}). The \textcolor{orange}{Unlabeled} group showed similar patterns (69\% AI for \textcolor{blue}{Calm}, 39\% for \textcolor{violet}{Upbeat}; $OR = 2.33, z = 1.94, p_adj = 0.157$), though this did not reach significance after correction (Hypotheses~\ref{h:unlabeled_cs},~\ref{h:unlabeled_ui}). No significant difference emerged between \textcolor{red}{Incorrect} and \textcolor{orange}{Unlabeled} groups ($p\_adj = 1.00$). Experimental group did not significantly predict efficacy ratings ($\chi^2(4) = 5.29, p = 0.259$). All groups rated human music as more effective overall (\textcolor{ForestGreen}{Correct}: 61\% human, \textcolor{red}{Incorrect}: 53\% human, \textcolor{orange}{Unlabeled}: 63\% human), with no significant pairwise differences (all $p\_adj > 0.72$) (Hypotheses~\ref{h:unlabeled_general},~\ref{h:labeled_general}).

\subsubsection{Impact of Emotion Condition}
Emotion condition significantly influenced preference ($\chi^2(2) = 9.40, p = 0.009$). Participants showed higher AI preference for \textcolor{blue}{Calm} music (60.5\%) compared to \textcolor{violet}{Upbeat} music (42.1\%; $OR = 0.61, z = -1.15, p = 0.249$). Within-group analyses revealed significant emotion effects for the \textcolor{ForestGreen}{Correct} ($\chi^2(2) = 8.42, p = 0.015$) and \textcolor{orange}{Unlabeled} groups ($\chi^2(2) = 9.09, p = 0.011$), with marginal effects in the \textcolor{red}{Incorrect} group ($\chi^2(2) = 5.18, p = 0.075$). The \textcolor{orange}{Unlabeled} group showed the most dramatic shift: 68.6\% AI preference for \textcolor{blue}{Calm} music dropped to 39.2\% for \textcolor{violet}{Upbeat} music (29.4\% decrease) (Hypotheses~\ref{h:unlabeled_ui},~\ref{h:unlabeled_cs},~\ref{h:labeled_ui},~\ref{h:labeled_cs}).

Emotion condition significantly influenced efficacy ratings ($\chi^2(2) = 6.40, p = 0.041$), with participants 58.7\% less likely to rate AI as emotionally effective for \textcolor{violet}{Upbeat} music ($OR = 0.41, z = -1.91, p = 0.056$). Within-group analyses revealed significant emotion effects for the \textcolor{ForestGreen}{Correct} group ($\chi^2(2) = 6.29, p = 0.043$) and highly significant effects for the \textcolor{orange}{Unlabeled} group ($\chi^2(2) = 18.75, p < 0.001$). AI efficacy ratings in the \textcolor{orange}{Unlabeled} group dropped from 51\% for \textcolor{blue}{Calm} to 11.8\% for \textcolor{violet}{Upbeat} music (39.2\% decrease) (Hypotheses~\ref{h:unlabeled_ui},~\ref{h:unlabeled_cs},~\ref{h:labeled_ui},~\ref{h:labeled_cs}).

\subsubsection{Preference-Efficacy Directional Consistency}
Preference and efficacy agreement was approximately 71\% overall, with no significant differences by experimental group ($\chi^2(2) = 0.44, p = 0.804$) or emotion ($\chi^2(1) = 0.43, p = 0.511$). However, critical asymmetries emerged in directional agreement. 89\% of participants who preferred human music also rated it more effective, demonstrating strong internal consistency. However, only 54.8\% of participants who preferred AI music rated it more effective (34.2\% difference). This asymmetry was most pronounced in the \textcolor{orange}{Unlabeled} group for \textcolor{violet}{Upbeat} music, where only 25\% of AI-preferrers also rated AI as more effective. 65\% (13/20) of participants who preferred AI-generated \textcolor{violet}{Upbeat} music rated human-composed music as more effective (OR = 0.027, p < 0.001). Significant consistency differences between AI and human preferrers emerged for \textcolor{violet}{Upbeat} music across all groups (\textcolor{ForestGreen}{Correct}: $p = 0.002$, \textcolor{red}{Incorrect}: $p = 0.041$, \textcolor{orange}{Unlabeled}: p < $0.001$).

\subsubsection{\textcolor{orange}{Unlabeled} Group Labeling Accuracy}
Emotion condition was the strongest predictor of labeling accuracy ($\chi^2(1) = 28.37, p < 0.001$), with participants performing substantially better on \textcolor{violet}{Upbeat} (56.9\% correct) than \textcolor{blue}{Calm} music (13.7\% correct; $\beta = 2.92, p < 0.001$). For \textcolor{blue}{Calm} music, 80.4\% of participants mislabeled the human song as AI, while 76.5\% mislabeled AI as human. Labeling accuracy significantly predicted preference for both \textcolor{blue}{Calm} (Fisher's exact test, p = 0.003) and \textcolor{violet}{Upbeat} music ($p = 0.009$): 79.1\% of incorrect labelers preferred \textcolor{blue}{Calm} AI music, compared to 14.3\% of correct labelers (61.9\% versus 24.1\% for \textcolor{violet}{Upbeat} music). Critically, 94.1\% (32/34) of participants who preferred AI-generated \textcolor{blue}{Calm} music and 63.2\% (12/19) of those who preferred \textcolor{violet}{Upbeat} AI mislabeled the tracks as human-composed, suggesting ``AI preference'' could reflect preference for music participants believed to be human-composed. Labeling accuracy showed a marginally significant relationship with efficacy ratings for \textcolor{blue}{Calm} music ($p = 0.108$) and no relationship for \textcolor{violet}{Upbeat} music ($p = 1.00$). Among participants who rated AI as more effective for \textcolor{blue}{Calm} music, 92\% (23/25) mislabeled the music as human. However, for \textcolor{violet}{Upbeat} music, only 50\% (3/6) of participants who thought AI as more effective mislabeled the song.

\subsubsection{Impact of Demographics} 
No demographic variables significantly predicted preference-efficacy agreement (all $p>0.1$), indicating that the preference-efficacy disconnect occurs consistently across demographic groups. Age significantly predicted AI preference ($\chi^2(7) = 17.42, p = 0.015$), with participants aged 34-41 showing dramatically elevated AI preference (78.7\%) compared to other age groups ($range: 34.6-66.7\%; \beta = 2.23, p = 0.014$). Emotion condition remained a strong predictor ($\chi^2(1) = 14.00, p < 0.001$), with significantly lower AI preference for \textcolor{violet}{Upbeat} music across demographics. Counterintuitively, prior AI experience ($\chi^2(4) = 5.31, p = 0.257$) and opinions about AI in music ($\chi^2(5) = 4.19, p = 0.523$) did not predict AI preference. Notably, participants who strongly opposed AI in music still preferred AI-generated music 66.7\% of the time, consistent with a mislabeling-driven preference pattern. Prior AI experience significantly predicted labeling accuracy ($\chi^2(3) = 16.94, p < 0.001$) but, interestingly, participants with average prior experience with AI were significantly worse at correctly identifying AI music ($\beta = -4.31, p = 0.018$). Age ($p = 0.201$) and musical experience ($p = 0.683$) did not improve labeling accuracy, suggesting that difficulty distinguishing AI from human music may transcend both technical familiarity and musical expertise.

\subsection{STAI}
Exploratory STAI-S results indicate that brief music exposure may impact momentary anxiety. Following the listening task, STAI-S scores decreased in 74 participants, increased in 50, and remained stable in 28, with clinically meaningful changes ($\ge8$ points) observed in both directions (Hypothesis~\ref{h:stai}). However, these results remain inconclusive. While we cannot isolate specific effects due to the pre-post design, these findings underscore the need for further research into the roles of trust, perceived authenticity, and listener context in shaping health outcomes.\footnote{Please see Supplementary Material~\ref{appendix:stai} for a more in depth discussion of STAI-S results}

\subsection{GEMIAC}
Estimated marginal means were calculated to examine differences in participants' emotional responses. In the \textcolor{blue}{Calm} condition, participants reported significantly higher ratings for low arousal/high valence emotions such as ``Relaxed, peaceful'' ($\beta=2.309, p<0.001, M = 3.32, SE = 0.068$), ``Full of tenderness, warmhearted'' ($\beta=1.510, p<0.001, M = 2.70, SE = 0.068$), and ``Moved, touched'' ($\beta=1.333, p<0.001, M = 2.48, SE = 0.068$) (Hypotheses~\ref{h:unlabeled_cs},~\ref{h:labeled_cs}). Conversely, the lowest-rated emotions in this condition were ``Agitated, aggressive'' ($M = 1.13, SE = 0.068$) and ``Tense, uneasy'' ($M = 1.17, SE = 0.068$), suggesting that calming music effectively minimizes arousing negative emotions. In contrast, in the \textcolor{violet}{Upbeat} case, participants reported significantly higher ratings for high arousal/high valence emotions, such as ``Energetic, lively'' ($\beta=1.118, p<0.001, M = 2.98, SE = 0.068$), ``Joyful, wanting to dance'' ($\beta=0.804, p=0.001, M = 2.66, SE = 0.068$), and ``Inspired, enthusiastic'' ($\beta=0.373, p=0.142, M = 2.51, SE = 0.068$) (Hypotheses~\ref{h:unlabeled_ui},~\ref{h:labeled_ui}). Notably, the lowest-rated emotions were ``Melancholic, sad'' ($M = 1.18, SE = 0.068$) and ``Agitated, aggressive'' ($M = 1.25, SE = 0.068$). Pairwise comparisons between GEMIAC categories revealed several significant contrasts. In the \textcolor{blue}{Calm} condition, ``Relaxed, peaceful'' was rated significantly higher than nearly all other emotions, particularly ``Tense, uneasy'' ($z = 29.29, p < 0.0001$). In the \textcolor{violet}{Upbeat} condition, ``Energetic, lively'' was significantly higher than ``Melancholic, sad'' ($z = 24.49, p < 0.0001$).

These findings suggest that music categorized as \textcolor{blue}{Calm} reliably elicited feelings of tranquility, tenderness, and awe, while \textcolor{violet}{Upbeat} music was associated with increased energy, enthusiasm, and movement. Overall, it appears that the specific GEMIAC \textit{categories} identified were not influenced by labeled origin, since human and AI music elicited similar responses. However, GEMIAC \textit{scores} showed some sensitivity to music origin. In the \textcolor{violet}{Upbeat} condition, the highest-rated GEMIAC category across conditions for both AI and human music was “Energetic, lively.” In the \textcolor{orange}{Unlabeled} ($p = 0.015$) and \textcolor{ForestGreen}{Correct} ($p = 0.0014$) groups, scores were significantly higher for human music than AI (Hypotheses~\ref{h:unlabeled_ui},~\ref{h:labeled_ui}). By contrast, no significant differences between scores for AI and human music were observed for “Relaxed, peaceful” in the \textcolor{blue}{Calm} condition (Hypotheses~\ref{h:unlabeled_cs},~\ref{h:labeled_cs}) or within the \textcolor{red}{Incorrect} group (Hypotheses~\ref{h:labeled_cs},~\ref{h:labeled_ui}). These results suggest listeners may find human-composed music more effective in eliciting emotional responses than AI-generated music in some circumstances, though further assessment is needed.

\subsubsection{Impact of Demographics}
Participants who reported excellent prior experience with generative music showed significantly higher overall GEMIAC scores ($\beta = 1.58, SE = 0.56, p = 0.005$), while other levels of experience (e.g., average, good, poor) were not significantly different from the reference, which could serve as an indication of confirmation bias due to prior positive exposure~\cite{klayman1995varieties}, though further investigation is needed. This effect was moderated by age, though age on its own did not yield significant effect. For example, those aged 34–41 with ``good'' prior AI music experience reported significantly higher ratings ($\beta = 1.963, SE = 0.895, p = 0.031$) than those aged 18–25 with no prior AI music experience. Additionally, participants who actively sought wellness music reported higher emotional responses ($\beta = 0.55, SE = 0.24, p = 0.026$) relative to those who did not. No significant main effects were found for general music experience or opinions on the role of AI in music. Post-hoc comparisons using the \texttt{emmeans} package confirmed robust differences in emotion intensity ratings between GEMIAC categories across both emotion case and participant-level covariates, suggesting that these emotional responses are consistent regardless of prior musical experience or AI attitudes. Overall, the results indicate that both participant backgrounds and specific emotional targets shape emotional impact. Random effects indicated modest variability across participants ($\sigma^2 = 0.24$), suggesting that individual differences contributed to emotional response patterns beyond fixed demographic predictors.  
\subsection{Free Response Data}
\subsubsection{Thematic Trends by Emotion Case}
In the \textcolor{blue}{Calm} case, participants frequently described music with references to ``peace,'' ``gentle[ness],'' and ``slower pace.'' \textcolor{ForestGreen}{Correct} participants often articulated clear emotional connections, describing music that ``takes me away to a more peaceful place'' or ``gave an overall calm feeling.'' In contrast, some \textcolor{red}{Incorrect} participants showed signs of cognitive dissonance, expressing confusion or second-guessing, though most indicated their preference/efficacy determinations did not change post-debrief, since they were shaped by personal taste, not origin labels. \textcolor{orange}{Unlabeled} participants often used generic descriptors like ``relaxing'' or ``not really different,'' potentially reflecting less confident appraisals.

In the \textcolor{violet}{Upbeat} case, participants focused on physical/energetic reactions, referencing tempo, rhythm, and movement (``danceable,'' ``energized,'' and ``joyful''). \textcolor{ForestGreen}{Correct} participants shared, ``It just made me feel better than the other song,'' and, ``[the human music] feels like an actual beat a person made.'' The \textcolor{red}{Incorrect} group often questioned whether high energy alone signaled human composition (``The [AI] song felt like it was trying too hard to make me feel something.'') \textcolor{orange}{Unlabeled} participants frequently cited technical details like beat and melody (``I couldn't tell a difference, but the rhythm stood out more.'')

\subsubsection{Thematic Frequencies and Cues}
From approximately 1,700 analyzed responses, we constructed nine main themes: \begin{itemize}
  \item Naturalness and Humanity (252 responses)
  \item Perceptual Ambiguity or Uncertainty (226)
  \item Emotional Resonance (225)
  \item Mechanical/Synthetic Quality (177)
  \item Technical Features (176)
  \item Listener Agency or Subjectivity (121)
  \item Genre or Context Association (54)\footnote{For analyses of themes with under 60 responses, please see Supplementary Material~\ref{appendix:continued_FRQ_analysis}.}
  \item Creativity, Novelty, or Usefulness(51)\footnotemark[\value{footnote}]
  \item Indifference or Detachment (50)\footnotemark[\value{footnote}]
\end{itemize}

\textbf{\textit{Naturalness and Humanity}} appeared most frequently in the \textcolor{ForestGreen}{Correct} and \textcolor{red}{Incorrect} groups, relating ``flow'', ``emotion'', ``realness'', ``naturalness'', ``organic quality'', ``soul'', and ``feel[ing] human'' to melody, beat, and rhythm. Interestingly, only one participant, in the \textcolor{ForestGreen}{Correct} group, specifically mentioned the impact of auditory artifacts, stating that ``the presence of more imperfections or audial artifacts'' would make human-composed \textcolor{violet}{Upbeat} music appear more AI. Interestingly, in the \textcolor{red}{Incorrect} and \textcolor{orange}{Unlabeled} groups, some assigned human characteristics to AI music, with one noting: \begin{quote}``...the human-composed piece [actually AI] blows the AI-generated piece [actually human] totally away, hands down. The AI piece [actually human] doesn't have nearly as much emotion...''\end{quote} \textbf{\textit{Mechanical/Synthetic Quality}} appeared frequently in the \textcolor{red}{Incorrect} group, possibly reflecting label-driven confusion. Phrases like ``too perfect'', ``robotic'', and ``artificial'' were common. One participant stated, ``it sounded like a fake performance by a human.'' \textbf{\textit{Perceptual Ambiguity or Uncertainty}} was most common in the \textcolor{red}{Incorrect} group. Participants often admitted confusion (``not sure'', ``couldn’t tell'', ``hard to say''). One explained, ``I would not have known it was AI unless you told me, it felt like a guess.'' Another noted, ``I had to just choose based on a hunch.'' In some cases, participants felt the pieces lacked distinguishing features. \textbf{\textit{Emotional Resonance}} appeared across groups via affective language such as ``calming,'' ``uplifting,'' ``enchanting,'' or ``joyful.'' While some noted, ``[they] just liked how it made [them] feel'', others included technical descriptors, indicating emotional interpretation coexists with analytical reasoning. \textbf{\textit{Technical Features}} were particularly mentioned in the \textcolor{orange}{Unlabeled} group, with focus on ``beat'' (``The beat felt natural...''), melody (``The melody...had a nicer flow.''), rhythm (`The rhythm helped build the emotion.''), tone (``The tone was softer...more organic.''), and tempo (``It had a slower tempo that helped me relax.''). Participants mentioned that they were, ``...not sure about the composer, but the instrumental was well layered.'' However, participants generally did not provide detailed insight into \textit{how} the specific elements they mentioned impacted preference or efficacy, echoing calls in prior work~\cite{yang2021examining} for further investigation into the impact of musical vocabulary on identification tasks. \textbf{\textit{Listener Agency or Subjectivity}} emphasized personal interpretation and taste (“for me”, “I think”, “depends on the person”), indicating awareness of subjectivity in musical evaluation. Participants situated their responses within personal frameworks rather than making universal claims (``It just felt better to me. I know others might disagree.”)

These findings raise broader questions about the evaluative frameworks used for determining preference and emotional efficacy. If listeners gravitate to perceived humanness or authenticity, this suggests social/relational dimensions remain central to emotion and value. Judgments guided by technical or aesthetic qualities indicate more evaluative or formal approaches. Understanding which features listeners prioritize helps explain variance in responses to AI versus human music. These results have implications for the design and training of future models and interfaces, which should account for the importance of human imperfection, emotional expressivity, and cultural context for meaningful emotional resonance. This insight is also critical for efforts to protect human musical creativity in an era where automated systems are increasingly capable of producing emotionally persuasive outputs.

\subsubsection{\textcolor{red}{Incorrect} Group Reflections on Mislabeling} 
Post-debrief, \textcolor{red}{Incorrect} participants reflected on whether their preference/efficacy designations changed. A chi-square goodness-of-fit test (Yes ($n = 17$), No ($n = 29$), and Unclear/Mixed ($n = 4$)) indicated participants were more likely to maintain their initial judgments ($\chi^2 = 18.76, p < 0.001$) despite the mislabeling, although many reflected on their assumptions or acknowledged surprising capabilities of AI. Additionally, many who reported a perspective change still emphasized the primacy of personal taste. Responses also included resistance to machine-authored art or a sense of validation regarding intuition. Individuals who felt Unclear/Mixed, however, typically showed partial reconsideration, such as surprise at songs' origin sources or acknowledgment of the quality of AI-generated music, without fully changing perspective.

\section{Discussion}
Using the GEMIAC, the STAI-S, and comparison tasks, we measured aesthetic emotional responses and momentary emotion regulation in reaction to AI-generated and human-composed music across \textcolor{ForestGreen}{Correct}ly labeled, \textcolor{red}{Incorrect}ly labeled, and \textcolor{orange}{Unlabeled} conditions for both \textcolor{blue}{Calm} and \textcolor{violet}{Upbeat} music. Our findings revealed a systematic preference-efficacy disconnect: participants tended to find human music more effective in eliciting target emotions but preferred AI music. This disconnect, driven in part by inability to accurately identify music origin, has critical implications for AI music system design, the role of perceived authenticity in emotional regulation, and the continued importance of preserving human musical creativity in an era of increasing automation. Though GEMIAC \textit{category} representation did not differ significantly by origin, GEMIAC \textit{scores} were significantly higher for human-composed music in specific cases, and qualitative data revealed persistent associations between perceived ``humanness'' and authenticity.

\subsection{Perceptions of Humanness and the Role of Labeling}
Exploratory thematic analysis emphasized qualities like ``flow'', ``realness'', ``organic[ness]'', ``soul'', and ``imperfection'' as indicators of humanity, though these patterns require validation through multi-rater coding. These descriptors echo prior work studying music performance that highlights the importance of micro-expressive timing, interpretive variability, and dynamic shaping to human musicality~\cite{chew2017performing, solomonova2023interpretive, repp1998variations, palmer1997music}. However, participants attributed human characteristics to AI-generated music when they believed it was human-composed. This pattern reveals a critical finding: perceived origin, rather than actual music origin, appeared to influence perception. In the \textcolor{orange}{Unlabeled} group, labeling accuracy was poor for \textcolor{blue}{Calm} music (13.7\% correct) but substantially better for \textcolor{violet}{Upbeat} music (56.9\% correct). For \textcolor{blue}{Calm} music, 80.4\% mislabeled the human song as AI and 76.5\% mislabeled AI as human, effectively reversing attributions. Most critically, 94.1\% of \textcolor{orange}{Unlabeled} participants who preferred AI-generated \textcolor{blue}{Calm} and 63.2\% who preferred AI in the \textcolor{violet}{Upbeat} cases mislabeled the music as human-composed. Participant labeling accuracy significantly predicted preference for both \textcolor{blue}{Calm} ($p = 0.003$) and \textcolor{violet}{Upbeat} music ($p = 0.009$). Among participants who mislabeled the music, 79.1\% preferred AI for \textcolor{blue}{Calm} compared to only 14.3\% among those who correctly labeled. These findings suggest observed ``AI preference'' reflected preference for music believed to be human-composed. Listeners also seemingly interpreted irregularities/flaws as hallmarks of authenticity. One participant remarked, ``...it felt like a real person was playing. There were tiny flaws that made it feel alive,'' suggesting listeners intuitively link imperfection with expressiveness. These findings underscore the importance of preserving traces of human imperfection while also affirming the ongoing cultural and emotional significance of human-made music.

\subsection{Preference, Efficacy, and Labeling Effects}
Three interrelated patterns emerged: (1) widespread mislabeling of AI music as human-composed, particularly for \textcolor{blue}{Calm} music; (2) preference ratings following participants' \textit{beliefs} about music origin; and (3) efficacy ratings more accurately reflecting \textit{actual} music origin, creating a systematic preference-efficacy disconnect. Overall, participants rated human-composed music as more effective (58.9\%) than AI-generated music (33.9\%), while preferences were more balanced (51.2\% AI vs. 43.1\% human). Multinomial logistic regression revealed participants were 52\% less likely to rate AI music as effective compared to their preference ratings ($OR = 0.48, p < 0.001$). Critically, this disconnect was asymmetric. Among participants who preferred human music, 89.0\% also rated it as more effective, demonstrating strong internal consistency. In contrast, only 54.8\% of AI-preferrers also rated it as more efficacious (a 34.2\% difference). This was most pronounced in the \textcolor{orange}{U} group for \textcolor{violet}{Upbeat} music, where only 25\% of AI-preferrers also rated AI as more effective. Remarkably, 65\% of participants who preferred AI-generated \textcolor{violet}{Upbeat} music rated human-composed music as more efficacious ($p < 0.001$). 

Experimental group significantly predicted preference ($\chi^2(4) = 16.51, p = 0.002$) but not efficacy ($\chi^2(4) = 5.29, p = 0.259$). The \textcolor{red}{Incorrect} group showed significantly higher AI preference (which they thought was human music) than the \textcolor{ForestGreen}{Correct} group ($p = 0.047$). All groups consistently rated human music as more effective, regardless of labeling condition and across demographic groups, suggesting that perceived origin drove the preference-efficacy asymmetry rather than musical qualities, age, AI experience, or musical expertise. Interestingly, participants with average prior experience with AI performed significantly worse at identifying AI music ($\beta = -4.31, p = 0.018$), and musical experience did not improve labeling accuracy ($p = 0.683$). Even participants who strongly opposed AI in music preferred AI-generated music 66.7\% of the time when they believed it was human-composed.

We identified a complex interplay between musical preference and emotional efficacy, since participants did not always prefer the music they found most emotionally effective. These results echo findings from Thompson (2006) where, during live performances of classical music, audiences were able to decouple perceived musical quality (which could be equated to the preference metric in this work) and emotion elicitation, ultimately suggesting that enjoyment was better predicted by emotional engagement than perceived quality~\cite{thompson2006audience,juslin2013everyday}. These findings challenge the use of preference as a primary metric in relaying user feedback. If participants prefer music that is not optimally emotionally effective, preference alone may not capture functional success for therapeutics or mood-modulation. This disconnect suggests aesthetic appreciation operates separately from perceived emotional effectiveness, with implications for how generative systems are evaluated, though further study is necessary.

\subsection{Design Implications}
Our results offer several actionable design insights for novel, emotionally intelligent musical interfaces that incorporate AI. Most notably, we find that listener preference does not always align with emotional efficacy. This challenges assumptions that preference sufficiently proxies affective success. Practically, these findings call for a deeper consideration of principles from music perception and psychology when designing novel musical interfaces. GEMIAC results demonstrate the need to calibrate system goals to the arousal level of the target emotion. In the \textcolor{blue}{Calm} condition, AI and human music were comparably effective in evoking low arousal/high valence emotions. However, in the \textcolor{violet}{Upbeat} condition, human music significantly outperformed AI on measures of "energetic, lively" in both \textcolor{orange}{Unlabeled} ($p = 0.015$) and \textcolor{ForestGreen}{Correct} ($p = 0.0014$) groups. This asymmetry suggests generative models may be more effective for low-arousal use cases (meditation, background music, wellness applications) while high-arousal contexts (motivation, celebration, live performance) necessitate human oversight or collaboration. Designers should carefully consider where AI-generated music is appropriate and ethically aligned, ensuring these tools augment rather than displace musicians.

Exploratory analyses of free response data suggest ``flow'', ``realness'', ``imperfection'', and ``soul'' signal human intentionality and creativity. Interestingly, participants used similar terms to described AI-generated music misidentified as human, indicating sonic cues can simulate human musicality. These findings highlight the enduring value of human expression and performance subtleties and caution against attempts to wholly replace/replicate them or strive solely for technical perfection.

Even if \textcolor{red}{Incorrect} group participants mentioned reevaluating the music after discovering true authorship, the majority did not change their preference/efficacy designations. This suggests that, while framing influences appraisal, emotional impressions are not entirely overwritten by authorship knowledge. Design strategies promoting transparency may balance openness with interpretive flexibility. Labeling should educate, helping listeners distinguish artistic collaboration from replacement, fostering support for human musicians. One such example is the \textit{Jordan and the jam\_bot} performance (2024), where Naseck et al. constructed a kinetic sculpture to help audiences understand whether musical output was human or AI-generated~\cite{naseck2025physical,blanchard2025jam_bot}. Many participants also emphasized the need for interpretive agency and subjectivity. Designers might support open-ended interpretation through interfaces that allow users to describe music's affective impact without requiring definitive labels, preserving space for personal meaning-making while pulling design principles from complimentary HCI work, including journaling tools~\cite{wang2025designing} .

\subsection{Limitations and Future Directions}
We note that this study is limited by stimulus variety and sample size. Free response data were coded by a single rater without inter-rater reliability assessment, limiting generalizability. While this work offers insights into listener interpretations of ambient and mood-based music, further research should use an expanded set of stimuli, exploring genre-specific reactions and cultural variability, given how participants with prior preference for wellness music reported higher GEMIAC emotional responses ($\beta = 0.55, SE = 0.24, p = .026$). Beyond this, even within our target genres, there are countless potential prompts and songs that could elicit completely different sets of labels, preferences, and efficacies. We also note that it is difficult to perfectly match stimuli on all possible covariates. However, we believe that this work serves as a meaningful first step. 

The lack of concrete discussion of auditory artifacts warrants future studies to assess the impact of musical and AI/ML expertise on emotional resonance and reflection. The preference-efficacy dichotomy deserves continued exploration through fine-grained musicological analysis and alternative assessment methods, such as confidence ratings and biometric assessment. Additionally, although STAI-S data were exploratory, they provide foundation for future work examining how human- versus AI-generated music affects anxiety and emotional regulation, potentially incorporating clinical populations, real-time physiological markers, or longitudinal exposure to evaluate therapeutic impact.

\subsection{Conclusion}
Our results call for a shift in how emotional generative music systems are evaluated and designed, challenging common metrics for evaluation and highlighting the need to distinguish aesthetic appreciation from functional outcomes. Instead of privileging surface-level preference or formal coherence, future systems should prioritize emotional credibility, interpretive openness, and relational trust. Designing to support, not supplant, human artistry is vital for affective outcomes and honoring musical experiences' social dimensions. Protecting space for human musicianship, emotional nuance, and creative labor must remain a central ethical commitment as generative music evolves.

\section{Ethical Standards}
This study met the ethical standards set by the Massachusetts Institute of Technology Committee on the Use of Humans as Experimental Subjects (MIT COUHES) protocol E-6149. This work was supported by discretionary funding from Opera of the Future at the MIT Media Lab. Author TRC was employed by Myndstream during the study period. To mitigate potential conflicts of interest, all data collection and analyses were conducted independently by KL in accordance with IRB-approved protocols.

\begin{acks}
Many thanks to Prof. Tod Machover and the Opera of the Future group; Freddie Moross, Adrienne O'Brien, Jordan Galvan, and the Myndstream team; and Dr. Anna Huang and the Human-AI Resonance Lab for their guidance and support. Thanks also to  Dr. Jin Ha Lee, Dr. William Brannon, Dr. Nikhil Singh, Stephen Brade, Jessie Mindel and Lancelot Blanchard for their feedback. 
\end{acks}

\bibliographystyle{ACM-Reference-Format}
\bibliography{references}

\clearpage
\appendix
 \section{Description of Musical Stimuli}
 \label{appendix:music}
 \subsection{AI-Generated \textcolor{blue}{Calm}}
 This AI-generated track is a 65-second instrumental composed using generative music techniques. The piece is characterized by its slow tempo (estimated at 72 BPM), sustained harmonic drones, and minimalistic structure. It features layered electric piano and synthesizer pads that blend into one another, creating a sense of stasis and sonic spaciousness. The track lacks the irregular fluctuations typically present in human performance, favoring smooth, consistent transitions between harmonic states. This design emphasizes emotional neutrality and ambient immersion, aligning with the genre conventions of ambient electronica or generative new-age music.

 \subsection{AI-Generated \textcolor{violet}{Upbeat}}
 This 65-second AI-generated track is designed to be upbeat and energizing, reflecting stylistic elements common in electronic dance or synthpop music. The tempo is estimated around 91 BPM, with clearly articulated rhythmic elements including synthesized drums and layered melodic loops. The track opens with a pulsing percussive pattern, underpinned by a steady synthetic bass line and accented by shimmering arpeggiated motifs. The harmonic content is relatively consonant, but the energy level is elevated compared to the calming version, with mid-range and high-frequency activity providing a bright, driving texture. The piece builds a sense of forward momentum through repetition and rhythmic layering, though it remains harmonically static with limited modulation or variation. The track's structure is loop-based and algorithmically regular, lacking the subtle dynamic shifts or expressive timing cues typical of human performance. It aims to produce a stimulating auditory experience through consistent rhythm, brightness, and moderate loudness.

 \subsection{Human-Composed \textcolor{blue}{Calm}}
 This human-composed relaxation track is a 72-second ambient instrumental that emphasizes softness, warmth, and expressive phrasing., the piece has a slow and unhurried tempo (approx. 53 BPM) and a fluid, rubato-like feel that allows for expressive timing deviations and natural breath. The harmonic progression unfolds gently, supported by resonant open strings and occasional melodic embellishments that arise within the arpeggiated figures. There is no accompanying percussion or rhythmic element, which places full focus on the keyboard’s tone and phrasing. The dynamic contour is subtle but present, with slight variations in volume and articulation contributing to an intimate, emotionally resonant performance. The piece evokes a quiet, introspective mood, and its recorded imperfections enhance the sense of authenticity and human presence, distinguishing it from its AI-generated counterparts.

 \subsection{Human-Composed \textcolor{violet}{Upbeat}}
 This human-composed invigorating piece is a 62-second upbeat instrumental that highlights rhythm and movement through natural performance. The primary instruments are electric guitar and synthesizer,played with an energetic strumming pattern and occasional syncopated accents. The estimated tempo is approximately 105 BPM, with consistent rhythmic drive provided by the interplay between percussive strumming and hand-played auxiliary percussion (such as a shaker or cajón). The piece is harmonically grounded in a major key, contributing to an overall positive and upbeat emotional tone. The recording captures the subtleties of human touch, including microtiming fluctuations, varied strumming intensity, and resonant overtones. Dynamic range is modest but perceptible, with occasional swells and decays adding expressive nuance. The spectral content reflects a balance between low-end rhythmic grounding and mid-to-high frequency harmonic richness, typical of live acoustic recordings. The overall impression is one of organic liveliness and human spontaneity, fitting within alternative and dance traditions.
 \subsection{Links to Audio Files}
 The human-composed and AI-generated music used in this study can be found at \url{https://web.media.mit.edu/~klecamwa/nime_2026_audio/} or at the QR code below:
 \begin{figure}[H]
 \includegraphics[width=0.4\columnwidth]{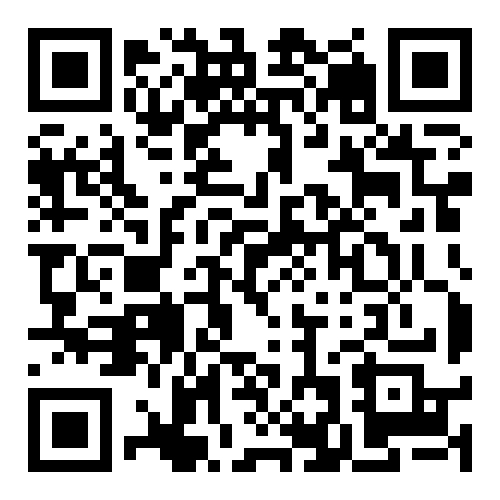}
   \Description{QR code linking to the human-composed and AI-generated music used in this study.}
   \label{fig:pref_eff}
 \end{figure}

 \section{GEMIAC Mock-Up}
 Please see Table~\ref{tab:gemiac}.
 \label{appendix:gemiac}
 \begin{table*}[]
 \begin{center}
 \begin{tabular}{c|c|c|c|c|c}
  & \textbf{Not at all} & \textbf{Somewhat} & \textbf{Moderately} & \textbf{Quite a lot} &\textbf{Very much}\\
 \hline
 Filled with wonder, amazed &1&2&3&4&5\\
 Moved, touched &1&2&3&4&5\\
 Enchanted, in awe &1&2&3&4&5\\
 Inspired, enthusiastic &1&2&3&4&5\\
 Energetic, lively &1&2&3&4&5\\
 Joyful, wanting to dance &1&2&3&4&5\\
 Powerful, strong &1&2&3&4&5\\
 Full of tenderness, warmhearted &1&2&3&4&5\\
 Relaxed, peaceful &1&2&3&4&5\\
 Melancholic, sad &1&2&3&4&5\\
 Nostalgic, sentimental &1&2&3&4&5\\
 Indifferent, bored &1&2&3&4&5\\
 Tense, uneasy &1&2&3&4&5\\
 Agitated, aggressive &1&2&3&4&5\\
 \end{tabular}
 \end{center}
 \caption{Mock-up of Coutinho and Scherer's ``Geneva Music-Induced Affect Checklist'' (GEMIAC) questions and rating scale~\cite{coutinho2017introducing}. Response prompt is as follows: \textit{Please rate the intensity (circle one) with which you experienced each of the following classes of feelings while listening to the piece of music on a scale ranging from 1 (not at all) to 5 (very much).}}
 \label{tab:gemiac}
 \end{table*}

 \section{Continued Free Response Analysis}
 \label{appendix:continued_FRQ_analysis}
 \textbf{Genre or Context Association} emerged in comparisons of the music to other functional or cultural reference points, including ads, video games, ambient background music, or specific genres. Such comparisons helped participants make inferences about authenticity or intention. One participant wrote that the \textcolor{violet}{Upbeat} AI-generated music, ``...sounded like it was made for an advertisement,'' implying both polish and emotional neutrality. These associations often shaped perceptions of purpose, complexity, or emotional depth. 

\textbf{Creativity, Novelty, or Usefulness} captured a small but distinct group of responses in which participants positively evaluated the music as ``interesting'', ``novel,'' or ``useful.'' These responses often reframed the AI–human distinction as less relevant than the generative potential of the technology. One participant described the AI-generated pieces as 11a tool for creativity'', suggesting that even synthetic-sounding outputs could have artistic value depending on context.

\textbf{Indifference or Detatchment} was expressed in participants that indicated that they ``...felt the same'' about all pieces, ``...felt indifferent'', or ``...didn’t care either way.'' Such responses may reflect low emotional impact, confusion, or lack of familiarity with music analysis, and tended to appear across conditions.

\section{Exploratory Insights from STAI and Implications for Wellness Music}
\label{appendix:stai}
Although the STAI-S data were collected for exploratory purposes, they provide early insight into how short-term music interventions may influence mood and anxiety. Overall, 74 participants’ STAI-S scores reduced after the study, 50 increased, and 28 were maintained. Before the study, 46 participants met the cutoff for indications of clinical anxiety, while 41 met the cutoff after. There was a significant reduction (a change of 8 points or more) in 21 participants, while 14 presented with a significant increase. With regards to the experimental groups, 25 in the \textcolor{ForestGreen}{Correct} class reduced their score and 16 increased; 24 in the \textcolor{red}{Incorrect} class reduced their score and 15 increased; and 24 in the \textcolor{orange}{Unlabeled} class reduced their score while 19 increased. While the causes of these shifts are not yet clear, several hypotheses merit future investigation. The reduction in anxiety for some participants may reflect acclimatization to the study environment, or a general calming effect from music engagement. Conversely, increases in anxiety in both emotion cases could reflect discomfort with the music itself, the framing of AI, or heightened self-monitoring due to the experimental context. Additionally, when examining the impact of music emotion case presentation order on STAI-S scores, we found no significant difference in score change based on whether participants ended with ``Upbeat and Invigorating'' (\textcolor{violet}{Upbeat}) or ``Calming and Soothing'' (\textcolor{blue}{Calm}) music ($p>0.05$).
 
We cannot, however, make statistical inferences from these data as there is no way to separate the impact of AI-generated music from human-composed music, since STAI-S data were collected before and after the full study, not each individual audio clip. As such, further investigation is necessary to assess significance and impact, though our exploratory results indicate promise in this approach. In light of these early findings, we plan to assess the therapeutic potential of both AI- and human-composed music in the frame of importance of listener perception and context. For wellness applications, the perceived authenticity, origin, and emotional resonance of music, as well as the level of trust users feel towards the intervention/stimulus, may critically shape outcomes. This study thus provides a foundation for future work exploring how trust and framing mediate the impact of music interfaces and interventions in healthcare, mindfulness, and public health contexts.
\end{document}